\documentclass[english,aps,prl,twocolumn,superscriptaddress]{revtex4-1}
\usepackage[english]{babel}
\usepackage{graphicx}
\usepackage[utf8]{inputenc}
\usepackage{amsmath}

\def\m{\mu} \def\n{\nu}

\newcommand{\be}{\begin{equation}}
\newcommand{\ee}{\end{equation}}
\newcommand{\bea}{\begin{eqnarray}}
\newcommand{\eea}{\end{eqnarray}}
\def\lab{\label} 
\def\tr{\textrm{tr}}

\DeclareMathOperator{\sign}{sign}
\begin{document}
\title{Holographic QCD in the Veneziano limit at finite Magnetic Field and Chemical Potential}
\author{Umut G\"ursoy}
\affiliation{Institute for Theoretical Physics and Center for Extreme Matter and Emergent Phenomena, Utrecht University, Leuvenlaan 4, 3584 CE Utrecht, The Netherlands}
\author{Matti J\"arvinen}
\affiliation{Institut de Physique Th\'eorique Philippe Meyer \& Laboratoire de Physique Th\'eorique, \'Ecole Normale Sup\'erieure, PSL Research University, CNRS, 24 rue Lhomond, 75231 Paris Cedex 05, France}
\author{Govert Nijs}
\affiliation{Institute for Theoretical Physics and Center for Extreme Matter and Emergent Phenomena, Utrecht University, Leuvenlaan 4, 3584 CE Utrecht, The Netherlands}
\begin{abstract}
We investigate the phase diagram of QCD-like gauge theories at strong coupling at finite magnetic field $B$, temperature $T$ and baryon chemical potential $\mu$ using the improved holographic QCD model including the full backreaction of the quarks in the plasma. In addition to the phase diagram we study the behavior of the quark condensate as a function of $T$, $B$ and $\mu$ and discuss the fate of (inverse) magnetic catalysis at finite $\mu$. 
In particular we observe that inverse magnetic catalysis exists only for small values of the baryon chemical potential. The speed of sound in this holographic quark-gluon plasma exhibits interesting dependence on the thermodynamic parameters.
\end{abstract}
\maketitle
\section{Introduction}

Despite being the established theory of one of the four fundamentals forces in Nature, the strong force, the phase diagram of quantum chromodynamics is still largely unknown. The elusiveness of the strong force is  the quarks and gluons becoming strongly coupled in the IR where most interesting dynamical phenomena, such as confinement and chiral symmetry breaking take place. In the presence of external magnetic fields, phenomena such as the magnetic catalysis \cite{Gusynin:1994re,Gusynin:1994xp,Gusynin:1994va} and the
recently discovered inverse magnetic catalysis \cite{Bali:2011uf,Bali:2012zg,D'Elia:2012tr,Ilgenfritz:2013ara} join this list. Interplay of these phenomena produces a rich phase diagram for QCD at finite temperature $T$, baryon chemical potential $\mu$, and magnetic field $B$ (see the review \cite{Andersen:2014xxa}). Understanding every corner of this phase diagram is crucial for multiple reasons that range from high energy physics, to astrophysics and cosmology. Indeed, the quark-gluon plasma produced in the heavy ion collision experiments and present at the core of neutron stars, magnetars, and in the early universe is believed to be strongly coupled and 
involve large magnetic fields \cite{Kharzeev:2007jp,Skokov:2009qp,Tuchin1,McLerran:2013hla,Vachaspati:1991nm,Tashiro:2012mf,Miransky:2015ava}. 
  
Lattice QCD, which, in fact can be viewed as a definition of QCD, and works also when the coupling is strong, has been an extremely fruitful method in the study of the phase diagram at finite temperature and magnetic field \cite{Bali:2011qj}. However this method is not fully functional \footnote{See however \cite{Aarts:2013uxa,Aarts:2015tyj} and the references therein for recent developments.} in the presence of baryon chemical potential due to the notorious {\em sign problem} \cite{Aarts:2015tyj}. This has prompted an investigation of QCD phase diagram at finite chemical potential using alternative non-perturbative methods, such as effective field theories \cite{Miransky:2015ava} and the holographic correspondence  \cite{Maldacena:1997re}. 

In this Letter we take the first step to explore the phase diagram at strong coupling using a full-fledged 
backreacted holographic model for QCD when the aforementioned thermodynamic parameters $T$, $\mu$ and $B$ are all finite. 
Namely, there is abundant literature on holographic methods employed in similar studies, see for example \cite{Ballon-Bayona:2017dvv} for the most recent entry in this list and the references therein.  Yet, most of this literature focuses on the limit of a large number of colors ($N_c$) and small number of flavors ($N_f$) where the effects of magnetic field on the system are suppressed due to the large imbalance between neutral particles, the gluons,  and charged particles, the quarks. To side-step this problem a realistic holographic effective theory where 
the fully backreacted 
contribution from the quark sector has been taken into account in the Veneziano limit \cite{Veneziano:1979ec}
\be\lab{Veneziano} 
N_c\to \infty, \qquad N_f\to \infty, \qquad x \equiv N_f/N_c = \textrm{const.}
\ee 
has been developed in \cite{jk}. Subsequently, a uniform external magnetic field in this model was introduced in \cite{Drwenski:2015sha}. What distinguishes our work from 
most of the other holographic approaches to QCD phase diagram 
lies here, that we consider the full contribution from the quarks. 

\begin{figure*}
\includegraphics[width=\textwidth]{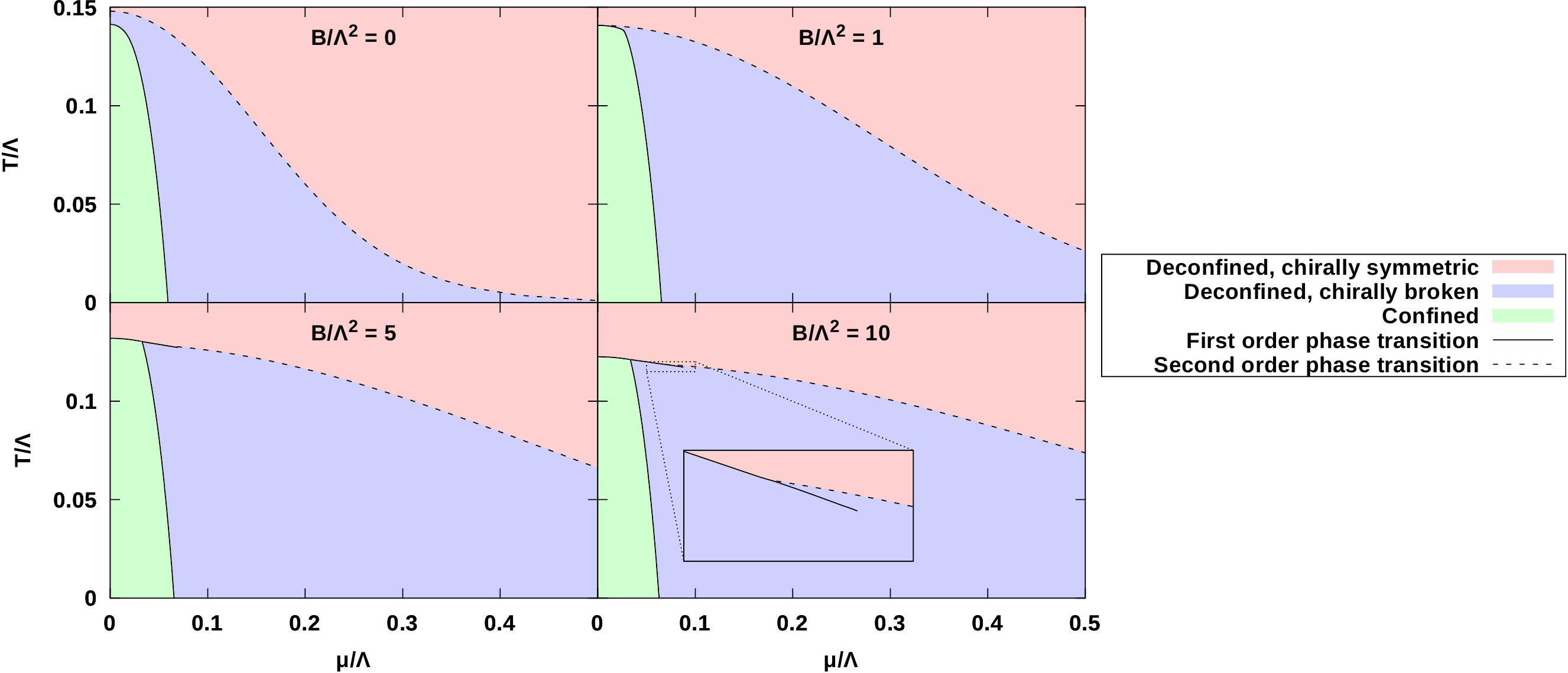}
\caption{\label{fig:phasediagram}The phase diagram on the $(\mu,T)$-plane at different values of the magnetic field. }
\end{figure*}

In the current paper,
in addition to finite $T$ and $B$, 
we extend the aforementioned model with finite chemical potential 
--- desired because this extension is currently very challenging to study by lattice techniques. The basic features of the model are reviewed in the next section. We then explore different corners of the resulting phase diagram in the following section. We uncover a rich structure with a first order 
confinement-deconfinement phase transition and a (typically) second order 
chiral phase transition, see figure \ref{fig:phasediagram}. We also observe that the sound speed in  
the plasmas in different phases exhibits a complicated dependence on $T$, $\mu$ and $B$.  

A central focus in our exploration is the {\em inverse magnetic catalysis} \cite{Bali:2011qj}, which is the weakening of the quark condensate, and consequently a decrease in the chiral transition temperature $T_\chi$, with increasing $B$. We stress that, as it has been demonstrated on the lattice \cite{Bruckmann:2013oba}, backreaction is essential in order to capture the dynamics of the inverse catalysis. This observation is also supported by the earlier holographic studies \cite{Jokela:2013qya,Dudal:2015wfn,Mamo:2015dea,Evans:2016jzo,Gursoy:2016ofp}.
A particularly pressing question is the fate of this phenomenon at finite $\mu$. We find that 
it is present in our model, also at finite $\mu$, albeit in a small range close to zero, as can be seen from figure \ref{fig:Tchi}.  As observed in \cite{Gursoy:2016ofp} and very recently discussed in detail in the case of the Sakai-Sugimoto model \cite{Sakai:2004cn} in \cite{Ballon-Bayona:2017dvv}, magnetization can be used to distinguish the magnetic and inverse magnetic catalysis. We also present our findings regarding the behavior of the quark condensate and magnetization in this section.  Finally, in the last section we summarize our results and provide an outlook.  

\section{Holographic QCD model}
\lab{sec::model}
We model the system of strongly coupled quarks and gluons in the large $N_c$ limit by a so-called ``bottom-up" model of holographic QCD 
 ---  a five dimensional gravitational system tuned by hand to reproduce the salient features of QCD in the IR \cite{ihqcd1, ihqcd2, Gubser:2008ny}. This model, which originally only described the glue sector of QCD, was extended to include the quark sector \cite{Bigazzi:2005md,ckp} in \cite{jk,aijk2} with number of flavors are also taken to be large in correlation with the number of colors as in (\ref{Veneziano}). Thus, the 5D gravitational action contains two parts, corresponding to the two sectors, glue and flavor: 
\[
S = S_g[g_{\m\n}, \phi]+ x\, S_f[g_{\m\n},\phi, \tau, L_\mu^a, R_\mu^a, V_\mu]\, ,
\]
where $x$ is the flavor to color ratio defined in (\ref{Veneziano}), which we fix as $x=1$ in this work. The gravitational action contains one 5D bulk field corresponding to 
the most important marginal or relevant operators of QCD 
up to spin-two. These are the metric for the stress tensor, the dilaton for the scalar glueball operator $\tr\, G^2$, a complex scalar $\tau$ for the quark condensate $\langle\bar qq\rangle$ and non-Abelian gauge fields $L_\mu^a$, $R_\mu^a$ and $V_\mu$ for the left and right chiral currents conserved under the symmetry $SU(N_f)_L\times SU(N_f)_R$ and the baryon number $U(1)_B$. We 
introduce \footnote{Since $V_\mu$ is dual to the baryon number, our $B$ has flavor independent couplings to matter and is identified as $e B_\mathrm{phys}$ up to an $\mathcal{O}(1)$ numerical coefficient, where $B_\mathrm{phys}$ is the physical magnetic field.} 
the baryon chemical potential $\mu$ and a uniform magnetic field $B$ in the $x_3$ direction through the bulk gauge field dual to this baryon number: 
\[
V_\mu = (\Phi(z),-x_2B/2,x_1B/2,0,0),
\]
where $z$ is the holographic direction, and the boundary value of the scalar potential gives the baryon chemical potential $\mu = \Phi(0)$,  \cite{Alho:2013hsa,Gursoy:2016ofp}. 
\begin{figure*}
\includegraphics[width=0.7\textwidth]{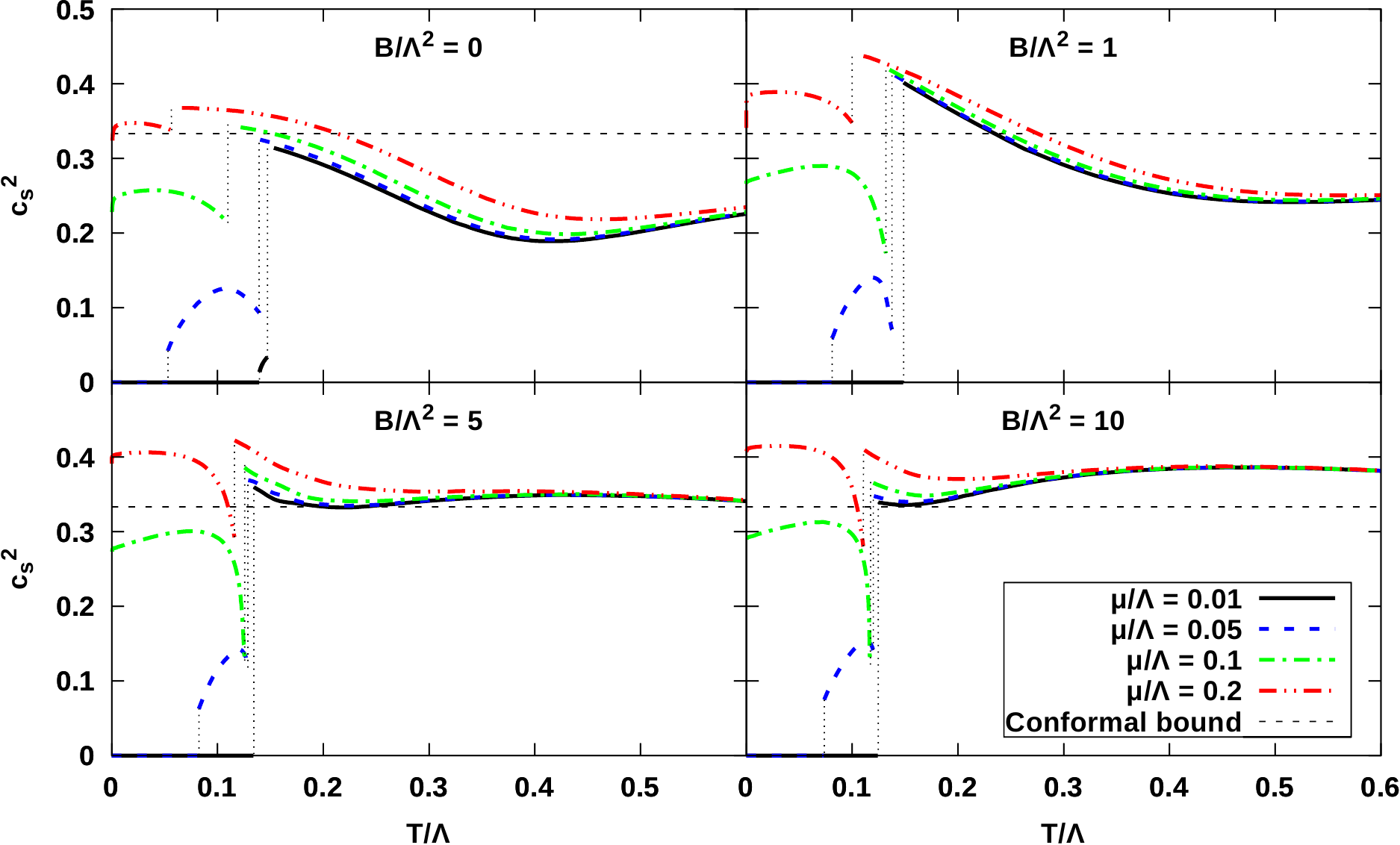}
\caption{\label{fig:cssq}The speed of sound squared $c_s^2$ as a function of temperature for different values of the magnetic field and the chemical potential. 
Numerical noise from these curves has been eliminated by using a high-momentum cutoff in Fourier space.}
\end{figure*}
The actions $S_g$ and $S_f$ above are taken precisely the same as in \cite{alte,Alho:2013hsa,Gursoy:2016ofp} with the potential parameter $c$ defined in \cite{Gursoy:2016ofp} fixed as $c=0.4$
so that the phase diagram at $\mu=0$ qualitatively agrees with lattice results. 
These actions are complicated and not illuminating, thus, we refer the interested reader to \cite{Alho:2013hsa,Gursoy:2016ofp} for details. It is worth mentioning that the model contains an energy scale $\Lambda$, that corresponds to the dynamically generated energy scale of QCD, which appears as an integration constant in the equations of motion.
Here we use this integration constant to define the dimensionless combinations  $T/\Lambda$, $\mu/\Lambda$ and $B/\Lambda^2$. The physical value of $\Lambda$ is very close to $1$~GeV.

\section{Phase diagram and sound speed} 
Thermodynamic properties of our holographic model follows from the holographic action evaluated on a given background solution. According to the AdS/CFT dictionary this corresponds to evaluating the free energy of the system in a given state. It is straightforward to check that the gravitational solutions satisfy the first law of thermodynamics $dF = -s dT - n d\mu - M dB$ where $s$, $n$ and $M$ are the entropy, baryon charge density and magnetization. In practice we calculate the free energy by first evaluating, $s$, $n$ and $M$ --- which is easier for a dual black-hole solution --- and then using the first law to integrate. An interesting phase diagram (at zero quark mass) results from competition between the following phases, as shown in figure \ref{fig:phasediagram}:  \\{\bf i)} a horizonless geometry with  a non-trivial profile for $\tau$, called the ``thermal gas" that corresponds to the chirally broken confined hadron gas, shown as green, \\ {\bf ii)} a black-hole solution with  a non-trivial profile for $\tau$, that corresponds to a deconfined quark-gluon plasma where the chiral symmetry is broken, shown as blue, \\{\bf iii)} a black-hole solution with trivial $\tau$, that corresponds to a deconfined quark-gluon plasma with restored chiral symmetry, shown as pink.

First, we notice that phase ii) which appears only in limited region of the phase diagram in the case $\mu=0$ of \cite{Gursoy:2016ofp}, extends into a sizeable part of the phase space at $\mu>0$. Second, we note that the deconfinement transition (between green and blue regions in the figure) is affected little by $B$ for smaller values of $T$. Essentially the effect of $B$ on the deconfinement transition is only significant when it merges with the chiral symmetry restoration for 
$B\gtrsim \Lambda^2$. We also observe that chiral symmetry restoration, becomes first order between $0<\mu/\Lambda\lesssim 0.1$ as $B$ grows. The first order line develops a second order endpoint and the second order transition branches off of the first order line (see inset in figure \ref{fig:phasediagram}).

Another thermodynamic observable that is very sensitive to the phase structure is the speed of sound, $c_s$ in the strongly interacting plasma. We study the speed of sound in the direction of the magnetic field, which can be computed by evaluating the derivative $-\mathrm{d}F/\mathrm{d}\epsilon$ keeping $n/s$ and $B$ fixed. One obtains,
\[
c_s^2 = \left.\frac{s\,\mathrm{d}T + n\,\mathrm{d}\mu}{T\,\mathrm{d}s + \mu\,\mathrm{d}n + B\,\mathrm{d}M}\right|_{n/s,B}\, .
\]
The result is shown in figure \ref{fig:cssq} as a function of $T$, $\mu$ and $B$. We observe that $c_s$ exhibits a jump precisely at the first and second order phase boundaries in figure \ref{fig:phasediagram}. We also find that it is enhanced by both $\mu$ and $B$ almost in the entire range of the parameter space. On the other hand, its dependence on $T$ is quite non-monotonic. As a side remark, we find that the conformal value of $c_s^2 = 1/3$ is crossed at various places somewhat unexpected from, but not in contradiction with, the findings of  \cite{Hohler:2009tv,Cherman:2009tw}. We checked that at the large $T$ --- not visible in figure \ref{fig:cssq} --- $c_s^2$ approaches the conformal value from below for all $\mu$ and $B$ considered in accordance with \cite{Hohler:2009tv,Cherman:2009tw}. 
\begin{figure}
\includegraphics[width=0.98\columnwidth]{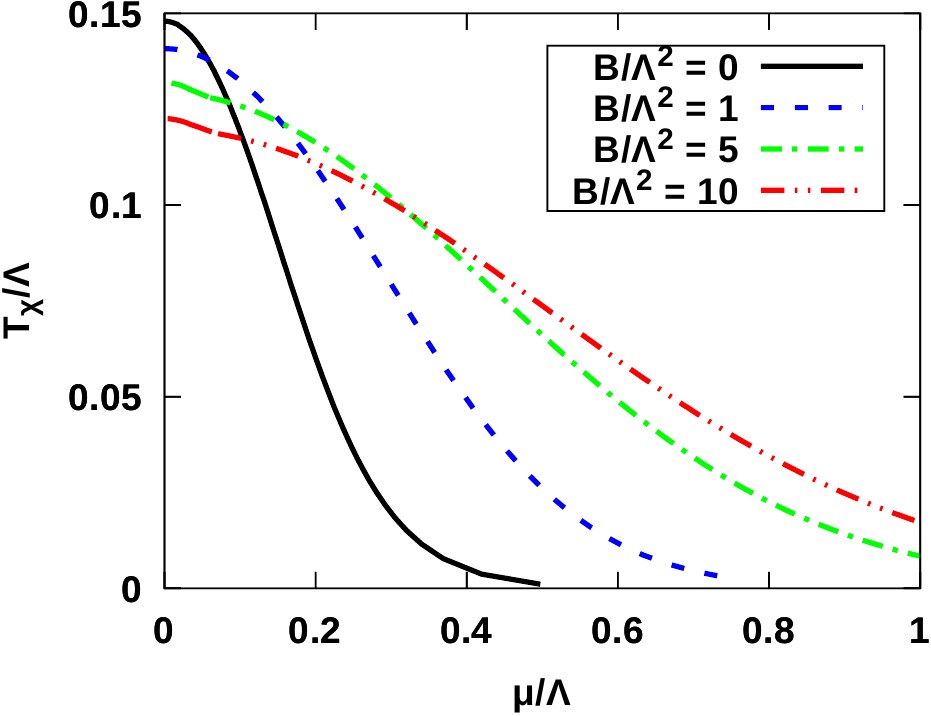}
\includegraphics[width=\columnwidth]{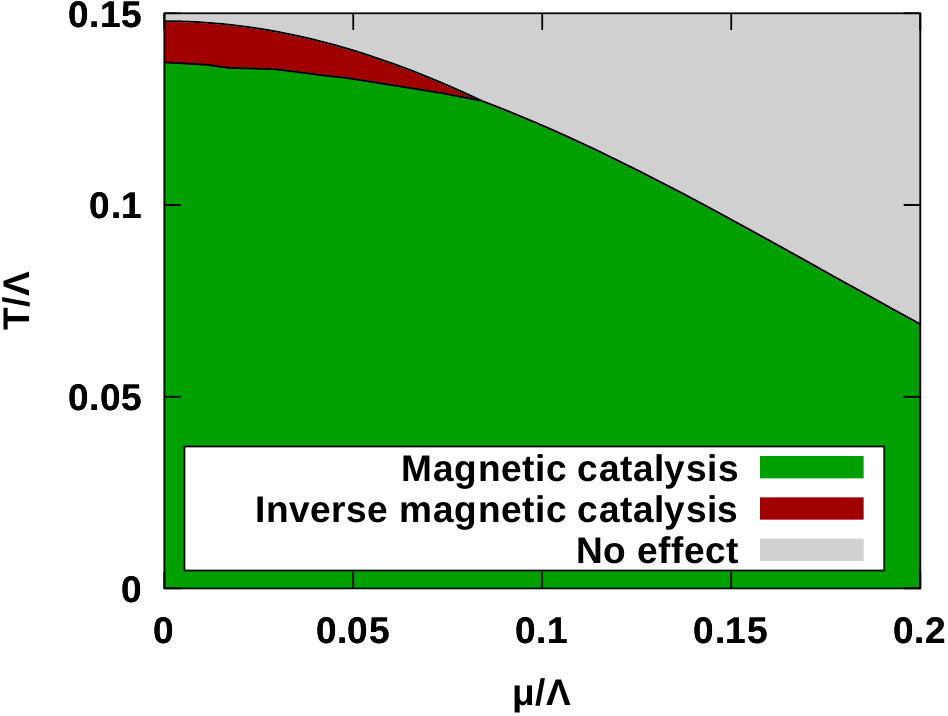}
\caption{{\em Top}: The chiral transition temperature $T_\chi$ as a function of chemical potential for different values of the magnetic field. Note that at large chemical potential, there is magnetic catalysis, while at small chemical potential, there is inverse magnetic catalysis. {\em Bottom}: Region where (inverse) magnetic catalysis occurs for small $B$, which is here defined by the sign of $\langle\bar qq\rangle_{B/\Lambda^2=0.1} - \langle\bar qq\rangle_{B/\Lambda^2=0}$\@.}
\label{fig:Tchi}
\end{figure}

\section{Inverse magnetic catalysis}
\lab{sec::IMC}
As discussed in the Introduction, a pressing issue is the dependence of the quark condensate on the magnetic field at finite chemical potential. One way to analyze this problem is to study the chiral transition temperature --- the phase boundary between the green or blue regions with the pink region in figure \ref{fig:phasediagram} --- as a function of $B$ and $\mu$ in more detail. In the figure \ref{fig:Tchi} (top), we observe that for sufficiently small $\mu$ the chiral transition temperature in fact decreases with $B$\@, indicating inverse magnetic catalysis. For larger values of $\mu$ magnetic catalysis takes over. In \ref{fig:Tchi} (bottom) we compare the regions of the phase space near small $B$ where these effects take place. We conclude that, increasing $\mu$ makes it harder for inverse magnetic catalysis to occur in general. 
Another observation from figure \ref{fig:Tchi} (top) is that the interval in $\mu$, where $T_\chi$ decreases with $B$, grows with increasing $B$. Therefore the region where inverse catalysis is found, which is somewhat limited in the case of small $B$ of figure \ref{fig:Tchi} (bottom), expands significantly as $B$ increases.
\begin{figure*}
\includegraphics[width=0.7\textwidth]{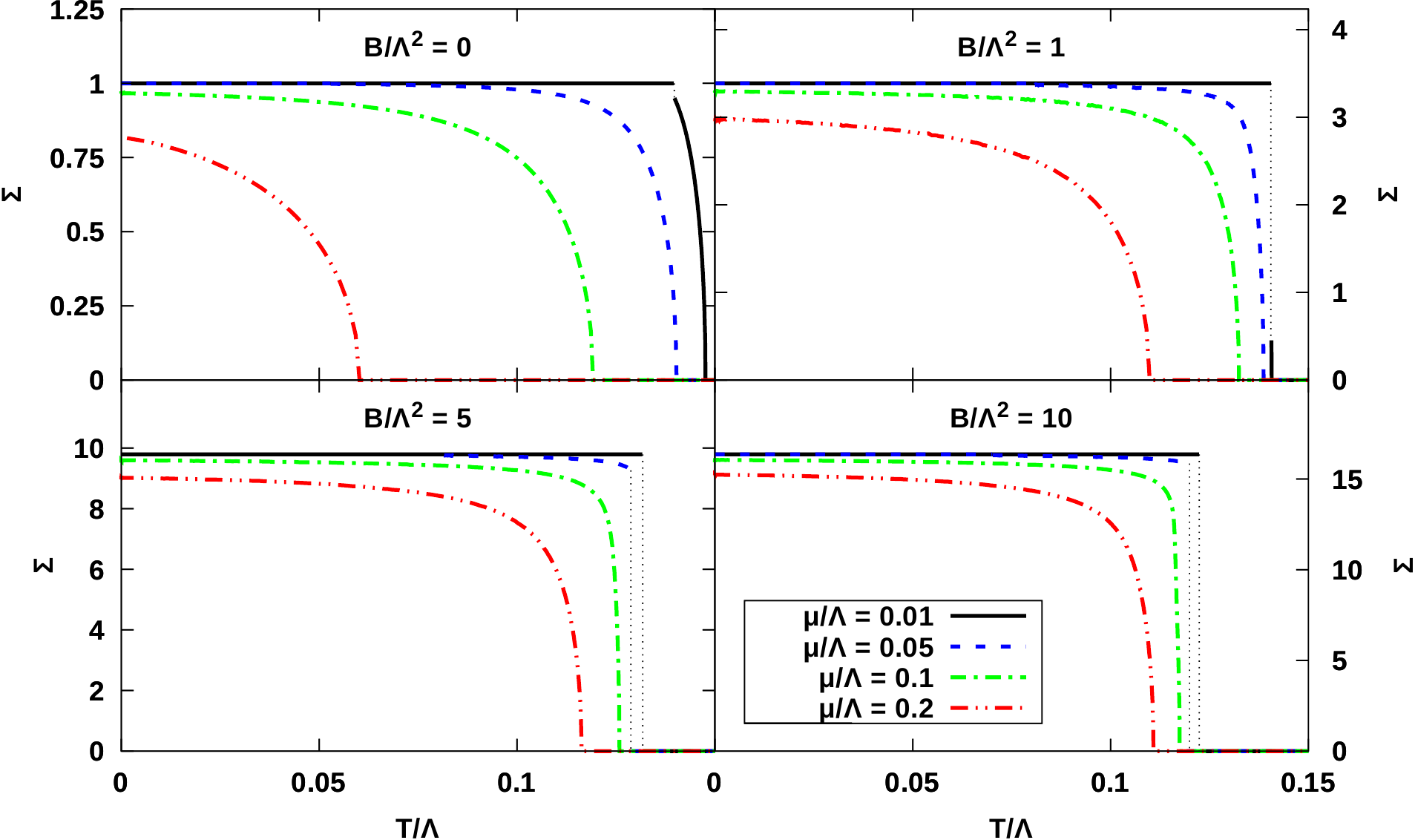}
\caption{\label{fig:qqbar}The normalized chiral condensate $\Sigma$ as a function of temperature for different values of the magnetic field and the chemical potential.}
\end{figure*}

We can also directly evaluate the quark condensate in our holographic model. As explained in \cite{Gursoy:2016ofp}, this can be read off from the near boundary asymptotics of the bulk complex scalar field $\tau$ that is dual to the quark condensate operator. Normalizing the condensate as 
\[
\Sigma(T,\mu,B) = \frac{\langle\bar qq\rangle(T,\mu,B)}{\langle\bar qq\rangle(0,0,0)}\, ,
\]
we plot in figure \ref{fig:qqbar} its dependence on $T$ for various choices of $\mu$ and $B$. We observe that it always decreases with $T$, making discontinuities at phase transitions in figure 
\ref{fig:phasediagram}. The reason for the absence of $T$ dependence of $\Sigma$ in the confined phase (green phase in figure \ref{fig:phasediagram}) is because this dependence is suppressed with $1/N_c^2$ in our model in the large $N$ limit \cite{Alho:2015zua,Gursoy:2016ofp}. One can also check that this figure is consistent with our observation of the inverse magnetic catalysis close to the chiral transition temperature. For example for $\mu = 0.05 \Lambda$ and $T = 0.12\Lambda$ the condensate attains a finite value at $B/\Lambda^2=5$ while it vanishes at $B/\Lambda^2 = 10$. 

\begin{figure*}
\includegraphics[width=0.7\textwidth]{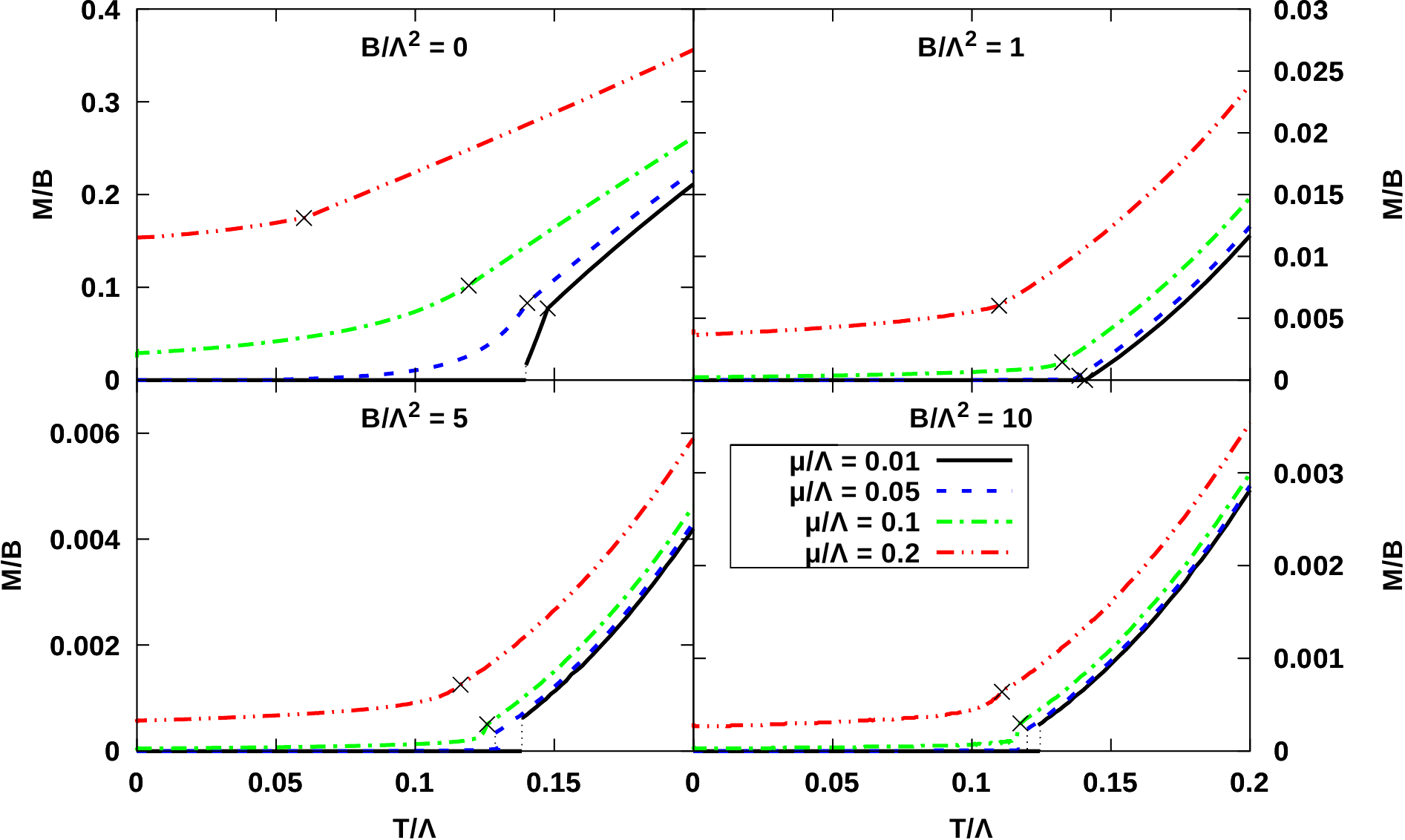}
\caption{\label{fig:magnetization}The magnetization divided by the magnetic field strength as a function of $T$, $\mu$, and $B$\@. The $B = 0$ result should be interpreted as the limit approaching $B = 0$\@.  The crosses denote the locations of the second-order chiral transitions. Here $M$ has been normalized so that $M_{T=0,\mu=0} = 0$ for every $B$\@.}
\end{figure*}
Finally, as observed in \cite{Gursoy:2016ofp} and \cite{Ballon-Bayona:2017dvv} and discussed in detail in the latter paper, magnetization can be utilized to distinguish the magnetic and the inverse-magnetic catalysis. In particular, whether the transition temperature increases or decreases with $B$ is correlated with whether the magnetization jumps up or down across a first order transition. Similarly for a second order phase transition, e.g. the dashed curves in figure \ref{fig:phasediagram}, one finds that
\[
\sign\left(\frac{\mathrm{d}T_\chi}{\mathrm{d}B}\right) = \sign\left(\frac{\mathrm{d}M(T_\chi + \epsilon)}{\mathrm{d}T} - \frac{\mathrm{d}M(T_\chi - \epsilon)}{\mathrm{d}T}\right),
\]
using the fact that the difference between the entropies $\Delta S(T_\chi(B),B) = 0$ for a second order transition. We observe in figure \ref{fig:magnetization} that, for example, for small $\mu$, $B$, one finds $\mathrm{d}T_\chi/\mathrm{d}B < 0$, an indication of inverse magnetic catalysis.
In general, although the kinks in $M(T)$ are often too small to be visible in figure \ref{fig:magnetization}, the findings agree with those in
figure \ref{fig:Tchi}.

\section{Discussion}

There are two main results in our paper. First is the phase diagram of a large-$N$ holographic QCD theory with full-backreaction from the quark sector at finite temperature, baryon chemical potential and magnetic field, cf. figure \ref{fig:phasediagram}. We considered massless quarks and fixed the flavor to color ratio, equation (\ref{Veneziano}) to be unity in this work. Generalization to massive quarks and study of the phase diagram at different values of $x$ are two immediate future directions. Also, we have disregarded the possibility of inhomogeneous phases in this work. Whether they can occur and compete with the phases we obtained here is an interesting question. We have checked the thermodynamic stability of all the phases shown in figure \ref{fig:phasediagram}. It is remarkable that the phase ii) (blue region in figure \ref{fig:phasediagram}), which is the deconfined plasma phase with broken chiral symmetry seems to be a universal prediction of a variety of holographic models \cite{Aharony:2006da,Mateos:2006nu,Kobayashi:2006sb,ckp,jk}. We find that this phase is also present at every finite $B$ and moreover it covers a larger part of the phase diagram for larger $B$. 

Our second main result is that inverse magnetic catalysis that is observed on the lattice simulations \cite{Bali:2011qj} at vanishing $\mu$ is also present at finite $\mu$. However, it is important to point out the differences in the definitions of this phenomenon in the literature. In our work, we follow the definition 
following the lattice findings \cite{Bali:2011qj}, that is, we define inverse magnetic catalysis as the weakening of the quark condensate, and related to this, the decrease in the chiral transition temperature $T_\chi$ with increasing $B$. We observe both of these effects. The original work of Preis et al. \cite{Preis:2010cq} and the most recent follow up \cite{Ballon-Bayona:2017dvv}, both using the Sakai-Sugimoto model \cite{Sakai:2004cn} define the phenomenon as the decrease in the critical chemical potential with $B$ for small temperatures. Interestingly, this latter effect has also been seen in approximations to QCD which are directly based on field theory (see, e.g., \cite{Andersen:2012bq}).
Notice also that we observe inverse catalysis at relatively low values of $\mu$, where the sign problem is probably surmountable. Therefore we expect that it will be possible to check our results on the lattice.

We do not dwell into the technicalities of our calculations in this Letter. Our calculations are based on the gravitational background described in \cite{Gursoy:2016ofp} in detail. This background is obtained by solving Einstein's equations numerically, and it is necessarily complicated since it involves backreaction of the flavor branes. To facilitate our calculations and minimize the error and time we wrote a program in C++, which was in part based on the earlier Mathematica code \footnote{T. Alho, \texttt{https://github.com/timoalho/VQCDThermo}.}. 
This program solves the equations of motion numerically using the LSODA algorithm, varying the boundary conditions in a grid to obtain data for the entire phase space. 
\section{Acknowledgments}

We are grateful to G. Aarts, J. O. Andersen, A. Ballon-Bayona, I. Iatrakis, A. Schmitt, and D. Zoakos for discussions. This work was supported, in part by the Netherlands Organisation for Scientific Research (NWO) under VIDI grant 680-47-518, and the Delta Institute for Theoretical Physics (D-ITP) funded by the Dutch Ministry of Education, Culture and Science (OCW).

\bibliographystyle{apsrev4-1}
\bibliography{mubpaper}
\end{document}